\documentclass[12pt]{article}
\usepackage{epsf, cite, amsmath, amssymb}
\usepackage{epsfig}

\makeatletter
\renewcommand\section{\@startsection {section}{1}{\z@}%
                                   {-3.5ex \@plus -1ex \@minus -.2ex}
                                   {2.3ex \@plus.2ex}%
                                   {\normalfont\large\bfseries}}
\renewcommand\subsection{\@startsection{subsection}{2}{\z@}%
                                     {-3.25ex\@plus -1ex \@minus -.2ex}%
                                     {1.5ex \@plus .2ex}%
                                     {\normalfont\bfseries}}
\makeatother

\let\non\nonumber

\newcommand{\bea}{\begin{eqnarray}}
\newcommand{\eea}{\end{eqnarray}}
\newcommand{\be}{\begin{equation}}
\newcommand{\ee}{\end{equation}}

\newcommand{\hlf}{\frac{1}{2}}

\newcommand{\R}{{\mathbb R}}
\newcommand{\M}{{\cal M}}

\newcommand{\mO}{\Omega}

\newcommand{\T}{\theta}

\newcommand{\e}{\epsilon}

\newcommand{\dd}{\delta}

\newcommand{\rr}{\rightarrow}
\newcommand{\m}{\mu}
\newcommand{\n}{\nu}
\newcommand{\p}{\partial}

\newcommand{\al}{\alpha}

\newcommand{\s}{\sigma}

\newcommand{\LRR}{\Longrightarrow}
\newcommand{\w}{\wedge}
\newcommand{\mo}{\omega}

\newcommand{\C}[1]{$(\ref{#1})$}

\begin{document}
\begin{titlepage}

\begin{center}

February 7, 2003
\hfill                  hep-th/0302049

\hfill EFI-02-44, SU-ITP-03/02

\vskip 2 cm
{\Large \bf Time-Dependent Warping, Fluxes, and NCYM}\\
\vskip 1.25 cm { Keshav Dasgupta$^{a}$\footnote{email: keshav@itp.stanford.edu}, Govindan
Rajesh$^{b}$\footnote{email: rajesh, robbins, sethi@theory.uchicago.edu},
Daniel Robbins$^{b 2}$, and Savdeep Sethi$^{b 2
}$
}\\
{\vskip 0.75cm
$^a$ Department of Physics, Stanford University, Stanford, CA 94305, USA \\
\vskip 0.2cm
$^b$ Enrico Fermi Institute, University of Chicago,
Chicago, IL
60637, USA\\}

\end{center}

\vskip 2 cm

\begin{abstract}
\baselineskip=18pt

We describe the supergravity solutions dual to D6-branes with both time-dependent
and time-independent $B$-fields. These backgrounds generalize the
Taub-NUT metric in two key ways: they have asymmetric warp factors and
background fluxes. In the time-dependent case,
the warping takes a novel form. Kaluza-Klein reduction
in these backgrounds is unusual, and we explore some of the new
features. In particular, we describe how a localized
gauge-field emerges with an analogue of the open string metric and
coupling. We also describe a gravitational analogue of the
Seiberg-Witten map. This provides
a framework in supergravity both for
studying non-commutative gauge theories, and for constructing novel
warped backgrounds.

\end{abstract}

\end{titlepage}

\pagestyle{plain}
\baselineskip=19pt

\section{Introduction}

The aim of this paper is to generalize the correspondence between
D6-branes and multi-Taub-NUT metrics~\cite{Townsend:1995kk}. The case
of particular interest to us is a D6-brane with NS-NS
$B_2$-fields along certain directions of its world-volume. 
The type IIA
geometry describing the back reaction of these branes has been studied
in~\cite{Alishahiha:1999ci, Alishahiha:2000qf}, and for massive type
IIA in~\cite{Singh:2002eu}. For a
particular choice of low-energy fields, the
world-volume theory is a supersymmetric non-commutative Yang-Mill
theory~\cite{Connes:1998cr}. On the other hand, the M theory
description of this background is a warped geometry with fluxes that
preserves one-half of the supersymmetries. For certain choices of
$B_2$-field, this background appeared
in~\cite{Chakravarty:2000qd}. Our approach differs from the AdS/CFT
style approach
of~\cite{Hashimoto:1999ut, Maldacena:1999mh}\  because we expect to see
a non-commutative gauge-field
directly in
our Kaluza-Klein spectrum. While it might be possible to see this
singlet field in the gravity dual of
lower-dimensional branes, the zero-mode analysis looks
harder~\cite{Adawi:1998ta}.

We
will also consider branes with time-dependent fluxes of the kind
described in~\cite{Hashimoto:2002nr}, and studied
in related work~\cite{Alishahiha:2002bk, Dolan:2002px, Cai:2002sv}. 
The world-volume theory is a
non-commutative gauge theory, but with a time-dependent
non-commutativity parameter~\cite{Hashimoto:2002nr}.
In these cases, the solution is warped in an unusual way not seen in
the string compactifications studied to date. We will see that the
metric for the {\it internal} space is also warped with a time-dependent
scale factor. A priori, it would have been hard to imagine that this
kind of solution could give rise to localized degrees of freedom. Yet the
existence of a brane dual suggests that this is indeed the case. 
The form of this solution also suggests the existence of a
far larger class of compactifications where the
internal space modulates in a time-dependent way. This direction will be 
investigated elsewhere~\cite{toappear}.

In the following section, we begin by obtaining the explicit
metrics and fluxes corresponding to the M theory duals of
D6-branes with time-independent fluxes. In section three, we repeat
this analysis for the case of D6-branes with time-dependent
fluxes. This gives us solutions with explicit time-dependent
warping. 

In the final section, we investigate the new issues
that arise in determining the
Kaluza-Klein spectrum for asymmetrically warped metrics of this
kind. In particular, we find that fluctuations of the metric and
$3$-form necessarily couple in these backgrounds. We show this in
two ways: first, by
a direct analysis of small fluctuations around the warped
background. Second, by a duality chasing argument. One of our main
goals in this investigation was to see how the non-commutative gauge
group arises from supergravity. This coupling of fluctuations makes it
clear that a
class of gravity gauge transformations constitute part of the gauge
theory symmetry
group.  We also explicitly find the harmonic 
$2$-forms on these spaces which play a key role in giving rise to space-time
gauge-fields. The couplings and metrics are deformed from the values
we expect for closed strings toward the values we expect for open
strings. 

While we see a good deal of evidence for the emergence of
non-commutative gauge symmetry in our Kaluza-Klein analysis, our
account from a direct small fluctuation analysis is still incomplete. Our duality
chasing argument suggests that the Seiberg-Witten map between
commutative and non-commutative variables map~\cite{Seiberg:1999vs} 
should follow naturally from T-duality! Indeed, we are able to obtain
this map from gravity to quadratic order in the gauge-fields. Our
analysis is really a natural extension of the argument by
Cornalba~\cite{Cornalba:1999ah}\ (see, also~\cite{Cornalba:1999hn, Ishibashi:1999vi})
to gravity. Using similar
reasoning, it should be possible to describe, in a uniform way, the coupling of the
supergravity multiplet to non-commutative gauge-fields. 
There is also a great deal more to be understood about
the time-dependent case. We should also point out that our
results have interesting implications for the geometric approach to
computing the NS 5-brane partition sum studied in~\cite{Dijkgraaf:2002ac, Bonelli:2000nz}. 
Lastly, we note that asymmetrically warped backgrounds have been
considered recently in string theory~\cite{Frey:2003jq}, and in non-stringy
settings~\cite{Csaki:2000dm}.

\section{Black Branes and T-duality}
\label{staticsolns}
\subsection{The smeared 5-brane solution}
\label{sm5}

Our starting point for most of the following discussion is the smeared black 5-brane
solution of IIB supergravity.  It can be obtained, for example, by T-dualizing
a spherically symmetric (in the transverse directions) 6-brane solution along
the brane.  The 5-brane solution is determined by the following dilaton,
$\Phi$, R-R potentials, $C_p$, and metric:
\bea
e^{2\Phi} &=& g_s^2 H^{-1} \non\\
ds^2 &=& H^{-\hlf}\left(-dx_0^2 + \cdots + dx_5^2 \right) + H^\hlf
\left(dx_6^2 + dr^2 + r^2 d\mO_2^2 \right) \non\\
C_6 &=& g_s^{-1}\frac{c}{r}H^{-1}dx_0 \w\ldots\w dx_5 \\
C_2 &=& g_s^{-1}c\cdot\cos\T dx_6 \w d\psi \non
\eea
where
\be
\label{sm5H}
H = 1 + \frac{c}{r}
\ee
and $c$ is a constant chosen to give the correct value for the 5-brane charge,
i.e. so that
\be
\frac{1}{(2\pi \ell_s)^2}\int_{\p\M_\perp} dC_2 =
\frac{1}{(2\pi \ell_s)^2}\int_{\p\M_\perp} \ast dC_6 = Q_5
\ee
where $\M_\perp$ is the space transverse to the brane (parametrized by $x_6$,
$r$, $\T$, and $\psi$).  For example, if $x_6$ is compact with radius $R$,
then $\p\M_\perp$ is $S_R^1 \times S_\infty^2 $, then we find
\be
\label{sm5c}
c = \hlf g_s \frac{\ell_s^2}{R}Q_5. 
\ee
In other cases this may be modified, as we shall see.  We will always
consider a single brane, so that $Q_5 = 1$.

We now proceed to generate some new solutions from this starting point.  Our
main tools are T-duality and lifting to M-theory. We outline our
conventions in Appendix \ref{conventions}.

\subsection{The usual story}

It is useful for us to begin with a brief discussion of the standard story
which we plan to generalize.  Typically, one starts with a spherically
symmetric 6-brane, but for practice,  we will instead T-dualize our smeared
5-brane along $x_6$.  If $x_6$ is taken to have radius $R$, we obtain:
\bea
\label{black6}
e^{2\Phi} &=& {\tilde g_s}^2 H^{-\frac{3}{2}} \non\\
ds^2 &=& H^{-\hlf}\left(-dx_0^2 + \cdots + dx_5^2 + dx_6^2 \right) + H^\hlf
\left(dr^2 + r^2 d\mO_2^2 \right) \non\\
C_7 &=& {\tilde g_s}^{-1}\frac{c}{r}H^{-1}dx_0 \w\ldots\w dx_5 \w dx_6 \\
C_1 &=& -{\tilde g_s}^{-1}c\cdot\cos\T d\psi \non
\eea
where $\tilde g_s = \ell_s g_s /R$ and $H$ is given by eqs. (\ref{sm5H}) and
(\ref{sm5c}).  Gratifyingly, it is the 6-brane solution that we expect.

Now lifting to 11D (with $\tilde g_s = 1$ for simplicity) we obtain:
\be\label{TN}
ds^2 = -dx_0^2 + \cdots + dx_6^2 + H\left(dr^2 + r^2 d\mO^2 \right) +
H^{-1}\left(dy + c\cdot\cos\T d\psi\right)^2. 
\ee
This solution corresponds to a KK monopole solution, with space-time
metric $\R^{1,6}\times\M$, where $\M$ is the Taub-NUT
manifold.

\subsection{The twisted compactification}
\label{twisted}

We will now generalize this solution to the case where the D6-brane
supports a rank 2 NS-NS $B_2$-field. Let us outline the steps:
our starting point is a D5-brane oriented along
$x_{0,1,2,3,4,5}$ and delocalised along $x_6$ (this is essentially T-dual
to a D6-brane oriented along $x_{0,1,2,3,4,5,6}$ and localised at a point 
in $x_{7,8,9}$). The directions $x_{5,6}$ form a square torus. We then 
{\it twist} the directions in such a way that a 
second T-duality along the
delocalised $x_6$ direction gives a D6-brane with a non-trivial $B_2$-field 
along $x_{5,6}$. We then lift this configuration to M theory where we
obtain a warped analogue of the Taub-NUT metric of \C{TN}\  with 4-form
$G_4$-fluxes. As a matter of notation, note that $G_4=dA_3$ where
$A_3$ is the $11$-dimensional supergravity potential. This will be the
solution that should give rise to a
non-commutative gauge-field. Let us give a detailed analysis of this procedure now.   

Starting with the solution from
section \ref{sm5}, we make
a change of coordinates (found in \cite{Chakravarty:2000qd}):
\bea
z_1 &=& \cos\al x_5 - \sin\al x_6 \non\\
z_2 &=& \frac{x_6}{R\cos\al}, 
\eea
or inverting,
\bea
\label{xtoz}
x_5 &=& \sec\al z_1 + R\sin\al z_2 \non\\
x_6 &=& R\cos\al z_2. 
\eea
We take $z_2$ to be compact with unit radius.  In these variables the
5-brane solution becomes
\bea
\label{tiltedD5}
ds^2 &=& H^{-\hlf}\left(-dx_0^2 + \cdots + dx_4^2 + \sec^2 \al dz_1^2
+ 2R\tan\al dz_1 dz_2 + R^2 \left(1 + \frac{c}{r}\cos^2 \al\right)dz_2^2
\right) \non\\
&& + H^\hlf \left(dr^2 + r^2 d\mO_2^2 \right) \\
C_6 &=& g_s^{-1}\frac{c}{r}H^{-1}dx_0 \w\ldots\w dx_4 \w\left(\sec\al dz_1 +
R\sin\al dz_2 \right) \non\\
C_2 &=& g_s^{-1}Rc\cdot\cos\al\cos\T dz_2 \w d\psi. \non
\eea
%
After T-dualizing along $z_2 $, we find
\bea
\label{rk2D6}
e^{2\Phi} &=& {\tilde g_s}^2 H_1^{-\hlf}H_2^{-1} \non\\
ds^2 &=& H_1^{-\hlf}\left(-dx_0^2 + \cdots dx_4^2 \right) +
H_1^\hlf H_2^{-1}\left(dz_1^2 + dz_2^2 \right) + H_1^\hlf \left(dr^2 +
r^2 d\mO_2^2 \right) \non\\
B_2 &=& H_2^{-1} \tan\al dz_1 \w dz_2 \non\\
C_7 &=& {\tilde g_s}^{-1}\frac{R'}{2r}H_2^{-1} dx_0 \w\ldots\w dx_4 \w
dz_1 \w dz_2 \\
C_5 &=& {\tilde g_s}^{-1}\frac{R'}{2r}H_1^{-1}\tan\al dx_0 \w\ldots\w dx_4
\non\\
C_3 &=& \hlf{\tilde g_s}^{-1}R'H_2^{-1}\tan\al\cos\T dz_1 \w dz_2 \w d\psi
\non\\
C_1 &=& -\hlf{\tilde g_s}^{-1}R'\cos\T d\psi \non
\eea
where we have defined
\bea
\label{rk2Hs}
\tilde g_s &=& \frac{\ell_s}{R}g_s \non\\
R' &=& \tilde{g}_s \ell_s \non \\ 
H_1 &=& 1 + \frac{R'}{2r\cos\al} \\
H_2 &=& 1 + \frac{R'\cos\al}{2r} \non
\eea
and have scaled $z_2 $ so that it has the natural T-dual radius of
$\ell_s^2 /R$.  The constant $c$ in (\ref{tiltedD5}) has been fixed in
(\ref{rk2D6}) and (\ref{rk2Hs}) by computing $\int dC_1$.

If we now set $\tilde g_s = 1$ and lift to M theory we obtain the solution:
\bea \label{ranktwoM}
ds^2 &=& H_1^{-\frac{1}{3}}H_2^\frac{1}{3} \left(-dx_0^2 + \cdots +
dx_4^2 \right) + H_1^\frac{2}{3} H_2^{-\frac{2}{3}}\left(dz_1^2 +
dz_2^2 \right) + H_1^\frac{2}{3} H_2^\frac{1}{3} \left(dr^2 +
r^2 d\mO_2^2 \right) \non\\
&& + H_1^{-\frac{1}{3}}H_2^{-\frac{2}{3}}\left(dy +
\hlf R'\cos\T d\psi\right)^2 \\
A_3 &=& H_2^{-1}\tan\al dz_1 \w dz_2 \w\left(dy + \hlf R'\cos\T
d\psi\right). 
\non
\eea
Note that the 3-form has a non-trivial field strength with $2$ legs in
the direction of $B_2$ and 2 legs along the internal space. 

Let us make the map between parameters explicit. Viewing the type IIA
configuration as our starting point, we began 
with a solution whose parameters are $\tilde g_s$, $\ell_s$, and the angle
$\al$ which determines the strength of the $B$-field via 
\be
B_2(r=\infty)-B_2(r=0) = \tan\al dz_1 \w dz_2. 
\ee
Our derivation of the IIA
supergravity solution led to a $B_2$ which vanished at the origin, but shifting
$B_2$ by any constant 2-form also satisfies the equations of motion.  When we
lift to 11 dimensions, we have new parameters $R_{11}$, $\ell_p$, and the
strengh of the 3-form, $A_3$.  In terms of the IIA parameters, the relation is
\bea
R_{11} &=& \tilde g_s \ell_s \non\\
\ell_p &=& \tilde g_s^{\frac{1}{3}} \ell_s \\
A_{z_1 z_2 y}(\infty) - A_{z_1 z_2 y}(0) &=& \tan\al. \non
\eea

\subsection{Generalizing to higher rank $B_2$-fields}

It is a simple matter to repeat the analysis above with additional twisted
compactifications, obtaining supergravity solutions corresponding to
$B_2$-fields of rank 4 or 6.

In the rank 4 case, one starts with a black 4-brane smeared in two directions
and performs two sets of coordinate redefinitions and T-dualities to obtain the
following black 6-brane configuration with flux:
\bea
e^{2\Phi} &=& {\tilde g_s}^2 H_0^\hlf H_1^{-1}H_2^{-1} \non\\
ds^2 &=& H_0^{-\hlf}\left(-dx_0^2 + dx_1^2 + dx_2^2 \right) +
H_0^\hlf H_1^{-1}\left(dz_3^2 + dz_4^2 \right) + H_0^\hlf H_2^{-1}\left(dz_5^2
+ dz_6^2 \right) \non\\
&& + H_0^\hlf \left(dr^2 + r^2 d\mO_2^2 \right) \non\\
B_2 &=& H_1^{-1} \tan\al_1 dz_3 \w dz_4 + H_2^{-1} \tan\al_2 dz_5 \w dz_6
\non\\
C_7 &=& {\tilde g_s}^{-1}\frac{R'}{2r}H_0 H_1^{-1}H_2^{-1} dx_0 \w dx_1 \w
dx_2 \w dz_3 \w dz_4 \w dz_5 \w dz_6 \\
C_5 &=& {\tilde g_s}^{-1}\frac{R'}{2r}H_1^{-1}\tan\al_2 dx_0 \w dx_1 \w dx_2
\w dz_3 \w dz_4 \non\\
&& + {\tilde g_s}^{-1}\frac{R'}{2r}H_2^{-1}\tan\al_1 dx_0 \w dx_1 \w dx_2 \w
dz_5 \w dz_6 \non\\
C_3 &=& -{\tilde g_s}^{-1}\frac{R'}{2r}H_0^{-1}\tan\al_1 \tan\al_2 dx_0 \w
dx_1 \w dx_2 \non\\
&& + \hlf{\tilde g_s}^{-1}R'H_1^{-1}\tan\al_1 \cos\T dz_3 \w dz_4 \w d\psi +
\hlf{\tilde g_s}^{-1}R'H_2^{-1}\tan\al_2 \cos\T dz_5 \w dz_6 \w
d\psi
\non\\
C_1 &=& -\hlf{\tilde g_s}^{-1}R'\cos\T d\psi. \non
\eea
We have defined
\bea
\tilde g_s &=& g_s \frac{\ell_s^2}{R_4 R_6} \non\\
H_0 &=& 1 + \frac{R'}{2r\cos\al_1 \cos\al_2} \\
H_1 &=& 1 + \frac{R'\cos\al_1}{2r\cos\al_2} \non\\
H_2 &=& 1 + \frac{R'\cos\al_2}{2r\cos\al_1}. \non
\eea
The lift to M theory has the form (again setting $\tilde g_s = 1$)
\bea
ds^2 &=& H_0^{-\frac{2}{3}}H_1^\frac{1}{3} H_2^\frac{1}{3} \left(-dx_0^2 +
dx_1^2 + dx_2^2 \right) + H_0^\frac{1}{3} H_1^{-\frac{2}{3}}H_2^\frac{1}{3}
\left(dz_3^2 + dz_4^2 \right) \non\\
&& + H_0^\frac{1}{3} H_1^\frac{1}{3} H_2^{-\frac{2}{3}}\left(dz_5^2 + dz_6^2
\right) + H_0^\frac{1}{3} H_1^\frac{1}{3} H_2^\frac{1}{3} \left(dr^2 +
r^2 d\mO_2^2 \right) \non\\
&& + H_0^\frac{1}{3} H_1^{-\frac{2}{3}}H_2^{-\frac{2}{3}}\left(dy +
\hlf R'\cos\T d\psi\right)^2 \\
A_3 &=& -\frac{R'}{2r}H_0^{-1}\tan\al_1 \tan\al_2 dx_0 \w dx_1 \w dx_2 \non\\
&& + \left(H_1^{-1}\tan\al_1 dz_3 \w dz_4 + H_2^{-1}\tan\al_2 dz_5 \w dz_6
\right)\w\left(dy + \hlf R'\cos\T d\psi\right). \non
\eea

Similar considerations can be applied to the rank 6 case.  We will
write down only the form of the 11D solution:
\bea
ds^2 &=& -H_0^{-1}H_1^\frac{1}{3} H_2^\frac{1}{3} H_3^\frac{1}{3} dx_0^2 +
H_1^{-\frac{2}{3}}H_2^\frac{1}{3} H_3^\frac{1}{3} \left(dz_1^2 + dz_2^2 \right)
+ H_1^\frac{1}{3} H_2^{-\frac{2}{3}}H_3^\frac{1}{3} \left(dz_3^2 + dz_4^2
\right) \non\\
&& + H_1^\frac{1}{3} H_2^\frac{1}{3} H_3^{-\frac{2}{3}} \left(dz_5^2 +
dz_6^2 \right) + H_1^\frac{1}{3} H_2^\frac{1}{3} H_3^\frac{1}{3} \left(dr^2
+ r^2 d\mO^2 \right) \non\\
&& + H_0 H_1^{-\frac{2}{3}}H_2^{-\frac{2}{3}}H_3^{-\frac{2}{3}}\left(dy +
\hlf R'\cos\T d\psi + \frac{R}{2r} H_0^{-1} t_1 t_2 t_3 dx_0 \right)^2 \\
A_3 &=& \frac{R}{2r} H_1^{-1} t_2 t_3  dx_0 \w dz_1 \w dz_2 + \hlf R H_1^{-1}
t_1 \cos\T dz_1 w\ dz_2 \w d\psi \non\\
&& + \frac{R}{2r} H_2^{-1} t_1 t_3  dx_0 \w dz_3 \w dz_4 + \hlf R H_2^{-1}
t_2 \cos\T dz_3 w\ dz_4 \w d\psi \non\\
&& + \frac{R}{2r} H_3^{-1} t_1 t_2  dx_0 \w dz_5 \w dz_6 + \hlf R H_3^{-1}
t_3 \cos\T dz_5 w\ dz_6 \w d\psi \non\\
\eea
with
\bea
H_0 &=& 1 + \frac{R'}{2rc_1 c_2 c_3} \non\\
H_1 &=& 1 + \frac{R'c_1}{2rc_2 c_3} \\
H_2 &=& 1 + \frac{R'c_2}{2rc_1 c_3} \non\\
H_3 &=& 1 + \frac{R'c_3}{2rc_1 c_2}. \non\\
\eea
and where $c_i \equiv \cos\al_i$ and $t_i \equiv \tan\al_i$.  Note the
appearance of a cross term in the metric between $dy$ and $dx_0$.  This is a
consequence of the fact that a D6-brane with a rank 6 $B_2$-field induces a
non-zero $C_1$ field (said differently, the D6-brane with a rank 6 $B_2$ carries
D0-brane charge).  This $C_1$ lifts to become the off-diagonal metric terms
seen above.

\section{Time-Dependent Cases}

Another new solution can be obtained by performing the null-brane
quotient~\cite{Figueroa-O'Farrill:2001nx, Simon:2002ma}\ on
the D5-brane, with the parabolic quotient acting along the brane (in directions $x^+$,
$x^-$, and $x = x_2$), and the shift direction transverse ($z = x^6$).  We
recall that the null-brane quotient acts by
\bea
&x^+ &\rr x^+ \non\\
&x^- &\rr x^- + 2\pi x + 2\pi^2 x^+ \non\\
&x &\rr x + 2\pi x^+ \\
&z &\rr z + 2\pi R. \non
\eea
Now let us switch to the natural invariant coordinates:
\bea
&\tilde x^- &= x^- - \frac{z}{R} x + \frac{z^2}{2R^2} x^+ \non\\
&\tilde x &= x - \frac{z}{R} x^+ \\
&\tilde z &= \frac{z}{R}. \non
\eea
In these coordinates, the quotient action is simply $\tilde z \rr \tilde z + 2\pi$.

In terms of these invariant coordinates (for simplicity, we will drop
the tildes from now on), the smeared 5-brane is given by
\bea
e^{2\Phi} &=& g_s H^{-1} \non\\
ds^2 &=& H^{-\hlf}\left(-2dx^+ dx^- - 2xdx^+ dz + dx^2 + 2x^+ dxdz +
(x^+)^2dz^2 + dx_3^2 + dx_4^2 + dx_5^2 \right)\non\\
&& + H^\hlf \left(R^2 dz^2 + dr^2 + r^2 d\mO_2^2 \right)\\
C_6 &=& -g_s^{-1}\frac{c}{r}H^{-1}dx^+ \w\left(dx^- \w dx + x^+ dx^- \w dz -
xdx\w dz\right)\w dx_3 \w dx_4 \w dx_5 \non\\
C_2 &=& g_s^{-1}Rc\cdot\cos\T dz \w d\psi, \non
\eea
where,
\be
H = 1 + \frac{c}{r},
\ee
and $c$ is the same as the case of an ordinary compact direction (see equation
(\ref{sm5c})).

Next, we T-dualize along $z$.  We obtain a solution for a D6-brane with
$B_2$ flux:
\bea
e^{2\Phi} &=& {\tilde g_s}^2 H^{-\hlf}h^{-1} \non\\
ds^2 &=& H^{-\hlf}\left(-2dx^+ dx^- + dx_3^2 + dx_4^2 + dx_5^2 \right)
+ H^\hlf h^{-1}\left(dx^2 + dz^2 \right) \non\\
&& + H^\hlf \left(dr^2 + r^2 d\mO_2^2 \right) +
H^{-\hlf}h^{-1}R^{-2}\left(-x^2 (dx^+)^2 + 2x^+ xdx^+ dx\right) \non\\
B_2 &=& h^{-1}R^{-1}\left(-xdx^+ + x^+ dx\right)\w dz \\
C_7 &=& -{\tilde g_s}^{-1}\frac{c}{r}h^{-1}dx^+ \w dx^- \w dx\w dx_3 \w
dx_4 \w dx_5 \w dz \non\\
C_5 &=& {\tilde g_s}^{-1}R^{-1}\frac{c}{r}H^{-1}dx^+ \w\left(x^+ dx^- -
xdx\right)\w dx_3 \w dx_4 \w dx_5 \non\\
C_3 &=& {\tilde g_s}^{-1}R^{-1}ch^{-1}\cos\T\left(-xdx^+ + x^+ dx\right)\w
dz\w d\psi \non\\
C_1 &=& -{\tilde g_s}^{-1}c\cdot\cos\T d\psi. \non
\eea
We have defined
\be
\tilde g_s = \frac{\ell_s}{R}g_s
\ee
which is the value of $\exp\left[\Phi(\infty)\right]$, and
\be
h = 1 + \frac{c}{r} + \left(\frac{x^+}{R}\right)^2
\ee
and where we have again rescaled $z$ so that it has the natural T-dual radius
$\ell_s^2 /R$.

Note that as $R\rr\infty$, we have $h\rr H$, $B_2 \rr 0$, and the solution
reduces to that of a standard spherically symmetric black 6-brane, as expected.

Finally, we would like to lift this configuration to M-theory.  We set
$\tilde g_s = 1$ to avoid cluttering the formulae:
\bea
\label{11Dtimedep}
ds^2 &=& H^{-\frac{1}{3}}h^\frac{1}{3} \left(-2dx^+ dx^- + dx_3^2 + dx_4^2
+ dx_5^2 \right) + H^\frac{2}{3} h^{-\frac{2}{3}}\left(dx^2 + dz^2 \right)
\non\\
&&  + H^{-\frac{1}{3}}h^{-\frac{2}{3}}R^{-2}\left(-x^2 (dx^+)^2 +
2x^+ xdx^+ dx\right)+ H^\frac{2}{3} h^\frac{1}{3} \left(dr^2 + r^2 d\T^2 +
r^2 \sin^2 \T d\psi^2 \right) \non\\
&& + H^{-\frac{1}{3}}h^{-\frac{2}{3}}\left(dy +
c\cdot\cos\T d\psi\right)^2 \\
A_3 &=& -R^{-1}h^{-1}\left(-xdx^+ + x^+ dx\right)\w dz\w\left(dy +
c\cdot\cos\T d\psi\right). \non
\eea
As a simple check on the algebra, we note that the equation of motion for
the M theory 3-form is obeyed; namely that $d\ast dA_3 = \hlf dA_3 \w
dA_3$ (both sides vanish in this case). This solution can be further
generalized by adding static $B_2$-fields along various directions of
the D6-brane. 

\section{Kaluza-Klein Reduction}
\subsection{Some preliminary comments}
\label{prelims}
The metrics and fluxes obtained in our prior discussion define both M
theory and type IIA compactifications. Our subsequent discussion
assumes an M theory compactification, but similar comments apply to type IIA. 
It is worth first recalling how we obtain localized 7-dimensional modes for
the case of vanilla Taub-NUT. We begin by considering small
fluctuations around our supergravity solution, 
$$ \left( 
\begin{array}{c}
\dd g \\  \dd A_3 \end{array}
\right), $$
where $\dd g$ parametrizes metric fluctuations, while $\dd A_3$
parametrizes 3-form fluctuations. In conventional situations without
background fluxes, the two
fluctuations decouple, and can be analyzed separately. 

So we consider 3-form fluctuations of the form~\cite{Imamura:1997ss},
\be
\dd A_3  = A_1 \wedge \omega + B_2 \wedge \xi , 
\ee  
with $A_1$ a 7-dimensional gauge-field, and $B_2$ a 10-dimensional
2-form. The normalizable closed
2-form $\omega$ is related to the one-form $\xi$ by the condition
$$ \omega = d \xi. $$ 
Despite appearances, $\omega$ is not trivial in cohomology because $\xi$ is not
normalizable. 
Roughly, reducing on $\xi$ gives us 10-dimensional propagating fields,
while reducing on $\omega$ gives 7-dimensional fields. The gauge
invariant field strength then takes the form, 
$$ d \dd A_3 = (F_2 + B_2)\wedge \omega, $$
which agrees with our expectations from string theory.  

The $U(1)$ gauge symmetry 
$$ A_1 \rr A_1 + d\lambda_0$$
visible at low-energies arises from the
symmetry
$$ A_3 \rr A_3 + d \lambda_2 $$
since  
$$ (A_1 + d\lambda_0) \wedge \omega = \dd A_3 + d
(\lambda_0 \omega). $$
Note that the gauge symmetry of the full theory 
is much larger since the gauged
7-dimensional Poincar\'e group is still unbroken by this
background. However, these two symmetry groups can be considered
separately in the low-energy theory. 

To proceed,  let us actually construct the 2-form $\mo$, which was found for Taub-NUT
in~\cite{Gauntlett:1996cw, Lee:1996if}. We will use the same approach
to find forms on our generalized Taub-NUT metrics.  The space of harmonic
two-forms can be decomposed into two components, each containing either self-dual
or anti-self-dual forms. Topologically, Taub-NUT is equivalent to $\R^4$ so any
closed 2-form $\mo$ is exact, and can be written in the form
$d\xi$. If $\mo$ is to be non-trivial then  $\xi$ cannot be normalizable.
Our search therefore reduces to finding
one-forms, $\xi$, satisfying $d\xi = \pm\ast_4 d\xi$.  To generalize our
discussion in a way that will be useful later, let us write our
Taub-NUT space in the form
\be\label{gtn}
ds^2 = G_1 (r)\left(dr^2 + r^2 d\mO_2^2 \right) + G_2 (r)\left(dy +
\beta\cdot\cos\T d\psi\right)^2. 
\ee
Here we assume that $y$ has periodicity $g_s \ell_s$, which means that
$\beta = \hlf g_s^{2/3} \ell_s$ for
the metric \C{gtn}. For this metric, we define vierbeins
\be
\begin{array}{cccc} e^r = G_1^\hlf dr, & e^\T = rG_1^\hlf d\T, &
e^\psi = r\sin\T G_1^\hlf d\psi, & e^y = G_2^\hlf \left(dy +
\beta\cdot\cos\T d\psi\right) \\\end{array}. 
\ee
We make the following ansatz for the form of $\xi$
\bea \label{xiansatz}
\xi &=& g(r)\left(dy + \beta\cdot\cos\T d\psi\right) \non\\
d\xi &=& g'(r)dr\w\left(dy + \beta\cdot\cos\T d\psi\right) -
\beta g(r)\sin\T d\T\w d\psi \\
&=& G_1^{-\hlf}G_2^{-\hlf}g'(r)e^r \w e^y -
\frac{\beta}{r^2}G_1^{-1}g(r)e^\T \w e^\psi \non\\
\ast d\xi &=& G_1^{-\hlf}G_2^{-\hlf}g'(r)e^\T \w e^\psi -
\frac{\beta}{r^2}G_1^{-1}g(r)e^r \w e^y. \non
\eea
{}For $\mo = d\xi$ to be SD (ASD), we require that $g(r)$ satisfy
\bea
g'(r) &=& \mp\frac{\beta}{r^2}G_1^{-\hlf}G_2^\hlf g(r) \\
&\LRR& g = \exp\left[\mp \beta\int^r G_1^{-\hlf} G_2^\hlf
  \frac{dr}{r^2}\right]. 
\non
\eea
To check normalizability, we integrate
\be
-\int\mo\w\mo = 2\beta\int g'(r)g(r)\sin\T dr\, d\T\, d\psi\, dy =
8\pi^2 \beta g_s \ell_s \left. \left[g(r)\right]^2 \right|_0^\infty. 
\ee
It will turn out that in all of the cases that we consider, the ASD solution
is normalizable, and the SD solution is not.  Also, we will find
that generally $g(0) = 0$, so the above formula reduces to
\be
\int \mo\w\ast\mo = 8\pi^2 \beta g_s \ell_s \left[g(\infty)\right]^2. 
\ee

In order to fix the normalization constant, we use the following argument
which appears in \cite{Sen:1997zb, Imamura:1997ss}. 
The action for a membrane wrapping the directions $r$, $y$, and a transverse
direction should give rise, on reduction along $y$, to the action of an open string
ending on the D6-brane.  The membrane action is
\be
S = \tau_{M2}\int A_3 = \frac{g_s^{-\frac{2}{3}}}{(2\pi)^2 \ell_p^3}\int
g'(r)dr\w dy \int A_1 = \frac{g(\infty)}{2\pi g_s^\frac{2}{3} \ell_s^2}
\int A_1
\ee
while the open string world-sheet action has a piece
\be
S = \int_{\p\Sigma} A_1. 
\ee
On comparing these two expressions, we find that $g(\infty) = 2\pi g_s^{2/3}\ell_s^2$, and so
\be
\label{genericg}
g(r) = 2\pi\ell_p^2 \exp\left[-\beta\int_r^\infty
G_1^{-\hlf}G_2^\hlf \frac{dr}{r^2}\right]. 
\ee

Returning to the case of standard Taub-NUT, we have $G_1 = H$ and
$G_2 = H^{-1}$.  The integral is particularly simple and gives,
\be
g(r) = 2\pi\ell_p^2 H^{-1}. 
\ee
Finally, let us see reduce 
part of the 11-dimensional SUGRA action using
$\mo$.  Ignoring $B_2$ for the moment, the kinetic term for the 4-form gives
\bea \label{TNcoupling}
S &=& -\hlf\frac{1}{2\kappa_{11}^2}\int d \dd A_3 \w\ast_{11} d \dd A_3 \non\\
&=& -\frac{1}{2(2\pi)^8 g_s^3 \ell_s^9} \, g_s^{-1} \, \int_{\R^{6,1}} dA_1
\w\ast_7 dA_1 \int_{\mathrm{TN}}\mo\w\ast_4 \mo \non\\
&=& -\frac{1}{2(2\pi)^4 g_s \ell_s^3}\int_{\R^{6,1}} dA_1 \w\ast_7
dA_1. 
\eea
This is the correct 7D YM action with the correct coupling constant, $g^2_{YM} \sim \ell_p^3.$

Let us imagine, for the moment, that we know the complete
7-dimensional effective action to all orders in $\ell_s$. We could now
contemplate moving in the space of SUGRA solutions by turning on a background
$<B_2> \neq 0$. By turning on this background, we reduce the full
$Spin(6,1)$ Lorentz group to some subgroup. 
Nevertheless, the low-energy physics should be captured by
the complete effective action which, from string theory, we expect takes the form 
\be \label{commDBI}
S_{eff} = \frac{1}{g_s (2\pi)^6 \ell_s^7} \int \, d^7x \, \sqrt{
  \det{ \left( {\bf 1} + 2\pi \ell_s^2 (F_2+B_2) \right) }} +
O(\partial F_2).    
\ee  
This is a completely commutative description of the low-energy
physics which has, among other features, linear couplings to the
background $B_2$.  This is
one way to describe the physics of our warped compactifications, but
it requires knowledge of physics beyond supergravity. We
now turn to a direct analysis of small fluctuations around the warped solutions.  

\subsection{The static warped case}
\label{staticwarped}
Reduction on warped metrics introduces a number of novel issues
to which we now turn. 
Let us begin by overviewing the key features of the supergravity
solutions described earlier.  The D6-brane world-volume
supports a $6+1$-dimensional abelian gauge-field. We turn on 
an NS-NS $B_2$-field  along certain directions of the
world-volume. The $B_2$-field
is characterized by its rank ($2$, $4$, or $6$). The presence
of the $B_2$-field explicitly breaks the $Spin(6,1)$ Lorentz symmetry to
a subgroup that depends on the rank. The corresponding M theory duals
are the warped metrics of section \ref{staticsolns}\ which generalize
the Taub-NUT space. 
By a warped metric, we mean
a metric that takes the form:
\be\label{usual} ds^2 = f(r) ds_{\rm space-time}^2 + ds_{\M}^2. \ee
The coordinate $r$ on
which the warp factor $f$ depends is along the compactification space
$\M$. We use the term
``compactification'' here in an abuse of terminology since $\M$ for us is a non-compact
manifold. Nevertheless, $\M$ supports normalizable modes which
propagate in space-time.

The $B_2$-field in type IIA lifts to the 3-form, $A_3$, with field
strength $G_4$ in M theory. The second important feature of these
solutions is the presence of $G_4$ flux. It is typical in
supergravity that warping is accompanied by fluxes. This
complicates a Kaluza-Klein analysis since the metric and $3$-form
modes can mix in a non-trivial way. A discussion of how to find
massless modes in situations like this appears
in~\cite{Dasgupta:1999ss}, which we will use as a guide. The first
change from the usual case of \C{usual}\ is that our warping is
asymmetric. Let us parametrize space-time by coordinates $x_0,
\ldots, x_4, z_1, z_2$, and let $B_2$, for simplicity, be non-vanishing in the $z_1,z_2$
directions. The metric takes the form, \be ds^2 = f_1(r) ds^2_{
\{x_0, x_3,x_4,x_5, x_6\} } + f_2(r) ds^2_{ \{z_1, z_2\} } +
ds_{\M}^2. \ee The accompanying M theory $G_4$ has $2$ legs in the
$z_1,z_2$ directions and $2$ legs in $\M$. We want to describe the
localized vector multiplets. More precisely, we want to
describe the leading terms in the action for a fluctuation $\dd
A_3$. The leading terms in the action are quadratic in the fluctuation
with all the
background parameters absorbed into the metric on the space of
fluctuations, 
$$ S_{eff} = \int d \dd A_3 \wedge \ast d \dd A_3 + \ldots. $$
This is quite different from the commutative description of
\C{commDBI} in which the background $<B_2>$ appears
explicitly even for the leading terms.
However, this is the usual procedure for determining the
effective action and light degrees of freedom around a given SUGRA
solution. This existence of (at least) two descriptions is very much
along the lines described  in~\cite{Seiberg:1999vs}. This approach should,
morally, give
the non-commutative description. If this is true then at least the
coupling constants and metric should be deformed toward the values we
expect for open strings. 

 In the unwarped case, vectors
arose by reducing $A_3$ on harmonic $2$-forms of $\M$. We need to be
more careful here. 
An $A_3$ fluctuation, $\dd A_3$, can be written in the form \be \label{usualexp}\dd A_3 =
\phi(x) C^{(3)} + A_1(x) \, C^{(2)} + \dd B_2 \, C^{(1)}\ee 
where $C^{(m)}$ is an $m$-form on the internal space. The
fields $\phi$ and $A_1$ have arbitrary dependence on $(x,z)$.
Since we want to consider vectors, let us set $\phi=0$ and $\dd B_2=0$.
Note that any vector $A_1$ is automatically part of a
supermultiplet that includes $3$ scalars. These additional scalars
come from metric fluctuations.
Now there is an immediate worry; namely, is $A_1$ a vector under
$Spin(6,1)$ or under $Spin(4,1)$? Since we have broken the symmetry
to $Spin(4,1)$ by an explicit $<B_{2}>$, it seems more natural to 
consider an expansion like
\be\label{generalansatz}  \dd A_3 = A_\m dx^\m \, C^{(2)}_1 + A_1 dz^1 \, C^{(2)}_2 + A_2 dz^2
\, C^{(2)}_3, \ee
where $\m =0,1,2,3,4$ and the $C^{(2)}_i$ are a priori independent. 
However, this decomposition does not seem natural if we
want to see a gauge symmetry in the effective theory that mixes the
$z_i$ and $x_\m$ directions. In this case,  for example, both
$A_\m dx^\m$ and $A_1 dz^1$ are needed to give a gauge covariant 
field strength, $F_{1\m} \,
dz^1 \wedge dz^\m$. Another possibility is to insist on an expansion
that involves just field strengths rather than potentials, but that
seems unnatural.  If we want an
expansion in terms of the supergravity potential $A_3$ rather than the field
strength $G_4$, it seems more natural to start by considering a fluctuation of the form
\be   \dd A_3 = \left( A_\m dx^\m  + A_1 dz^1 + A_2 dz^2 \right)
\wedge \omega, \label{fluct}\ee
where we introduce one internal 2-form $\omega$. We take this choice
as our starting point, although we will see in
section \ref{staticchase}\ that the more general ansatz of \C{generalansatz}\ is
actually possible. 

The fluctation is expanded in eigenmodes of the equation of motion
\be \label{eomC} d\,
\hat{*}\, d\, \dd A_3 = - G_4 \wedge d \dd A_3, \ee where $\hat{*}$
denotes the Hodge dual with respect to the warped metric.  
The right hand side of \C{eomC}, which comes from the
Chern-Simons interaction
$$ \int G_4 \wedge G_4 \wedge A_3, $$
in M theory, is
a $(4,4)$ form where $(p,q)$ denotes a $p$ form in space-time and
a $q$ form on $\M$. Using \C{fluct}\ which is a $(1,2)$ form, we see
that the left hand side of \C{eomC}\ never gives a $(4,4)$ form
so these terms decouple initially. 

The left hand side can be expanded to give, \bea \label{expansion}
d \, \hat{*}\, d
\dd A_3 &=& d* F_2 \, \wedge (f_1^{a+1/2} f_2^{1-a} \,* \omega) + *F_2
\wedge d (f_1^{a+1/2} f_2^{1-a} \, * \omega ) \cr & & + d*A_1\wedge
( f_1^{b+3/2} f_2^{1-b}* d \omega) 
+ *A_1\wedge d(f_1^{b+3/2} f_2^{1-b} * d\omega).\eea 
The
Hodge star products are now with respect to the unwarped
space-time metric and the metric for $\M$. The numbers $a,b$
depend on the number of legs that $F_2$ and $A_1$, respectively, have
in the directions of $<B_2>$. To make the last two terms vanish, we
require that $d\omega =0.$ 

The first term of \C{expansion}\ gives the equation of motion for
$A_1$, and it is
already clear that the metric and couplings will be asymmetric. We can
see this explicitly. Let us assume, for the moment, that the only
way gauge-field kinetic terms arise is from this first term. The
analogue of \C{TNcoupling}\ now gives a matrix of coupling constants,
\be \label{warpcoup} \int \omega \wedge (f_1^{a+1/2} f_2^{1-a} \,
*\omega). \ee
The value of the coupling now depends on which component of $F_2$ we
consider through the value of $a=0,1,$ or $2$. This is a feature
forced on us by the asymmetric
warp factors. For small $B_2$, we can evaluate \C{warpcoup}\ for a
harmonic $\omega$. The harmonic form on our internal space for the
rank $2$ $B_2$-field is
determined by using the same ansatz as in \C{xiansatz}. The function
$g(r)$ is again given
by \C{genericg}\ but
with different warp factors.  The result in this case (with the
correct normalization) is
\be
g(r) = 2\pi\ell_p^2 \left(1 + \cos\al\right)\left(1 +
\frac{1}{\cos\al}\right)\frac{2r}{\left(\sqrt{2r\cos\al + R'} +
\sqrt{\frac{2r}{\cos\al} + R'}\right)^2}. 
\ee
Using the explicit form for $\omega$, we see that
\be
S = \frac{1}{4g_{YM}^2}\sum_{a=0}^2 f^{(a)}(B)\int_{\R^7} F^{(a)} \w\ast 
F^{(a)}.
\ee
The index $a$ on $F$ again refers to how many legs the field strength has lying along
the $B_2$-field (e.g. $F^{(2)} \sim dz_1 \w dz_2$).  The coupling
$g_{YM}^2$ is the usual 
value for $B=0$ so all of the $B$-dependence is absorbed into the functions 
$f^{(a)}.$

Explicitly we find, in the limit of small $B$ that:
\bea \label{fa}
f^{(0)} &=& 1 + \frac{B^2}{6} - \frac{B^4}{16} + {\cal O}(B^6) \non\\
f^{(1)} &=& 1 - \frac{B^2}{6} + \frac{5B^4}{48} + {\cal O}(B^6) \\
f^{(2)} &=& 1 - \frac{B^2}{2} + \frac{7B^4}{16} + {\cal O}(B^6). \non
\eea
We now demand that our effective space-time action be covariant taking
the form
\be \label{cov}
S \sim \frac{1}{G_s}\int_{\R^7} \sqrt{G} G^{\alpha\beta}G^{\gamma\delta}F_{\alpha\gamma}
F_{\beta\delta}
\ee
for some coupling $G_s$ and some metric $G$ which can both
depend on $B$. Note that the indices $\alpha, \beta, \gamma,\delta$ in
\C{cov}\  run over all
$7$ space-time coordinates. If we assume that the $B$-dependence is of the form we
expect from open string physics~\cite{Seiberg:1999vs}
\bea
G_s &=& g_s u(B) \\
G_{\alpha\beta} dx^\alpha dx^\beta &=& dx^\m dx^\m + v(B)(dz_1^2 + dz_2^2)
\eea
with $u(0) = v(0) = 1$, then $G^{zz} = v(B)^{-1}$, 
$\sqrt{G} = v(B)$, then we expect the functional dependence of the
action \C{cov}\ on $B$ to be
\be
S \sim \frac{1}{4g_{YM}^2} u(B)^{-1} \int\left[v(B) F_{xx}F_{xx} + 
F_{xz}F_{xz} + v(B)^{-1}F_{zz}F_{zz}\right]. 
\ee
We can solve for these parameters in terms of the $f^{(a)}$ above
\bea
u(B) &=& (f^{(1)}(B))^{-1} \\
v(B) &=& \frac{f^{(0)}(B)}{f^{(1)}(B)} =
\frac{f^{(1)}(B)}{f^{(2)}(B)}. 
\eea
The second of these equations imposes a consistency check on our
solution. Indeed, if we expand our functions to arbitrary order in $B$
\be
f^{(a)}(B) = 1 + \sum_{i=1}^\infty c_i^{(a)} B^{2i}
\ee
then to order $B^2$,  the consistency check is simply that 
$c_1^{(0)}+c_1^{(2)} = 2c_1^{(1)}$, which is satisfied by our
solution \C{fa}. However, at order $B^4$, the functions $f^{(a)}$ 
of \C{fa}\ do not satisfy this constraint. This is not surprising for two reasons:
first, higher derivative corrections to supergravity like terms of the
schematic form $ \int || G_4||^8 $ can, in principal, 
contribute at $O(B^4)$. Second, as we shall now see, there are many
additional contributions to the gauge-field kinetic terms even at the
level of supergravity. 

To see this, let us return to our analysis of equation \C{expansion}. 
More
interesting, and problematic, than the first term is the second
term. This term is a $(5,3)$ form, and so
does not give a non-vanishing term in the action when wedged with $\dd
A_3$. However, it does give couplings between the supergravity fields
and the fields localized on the brane in a way described
in~\cite{Imamura:1997ss,Bergshoeff:1997gy}. 
For example, a term on the brane of the form
$$ \int \dd B_2 \wedge *F $$
arises this way. 
Usually, without warping, we would
pick a gauge where $\omega$ is harmonic
\be d * \omega = 0 \ee
to make this term vanish. This still works in the commutative
directions where $a=0$ if we choose the gauge,
\be \label{gaugezero}
d (* f_1^{1/2} f_2\,  \omega) =0,
\ee
but not for the non-commutative terms with $a=1,2$.\footnote{There is
  some freedom here to choose a gauge so that the problematic terms
  with $a=1$ or $a=2$ vanish rather than $a=0$. However, the problematic $a=0$ terms
  cannot be cancelled by a metric fluctuation in the way described
  subsequently. Rather, if these gauge choices are sensible then the $a=0$
  term must cancel against a new induced $\dd A_3$ fluctuation.}

 How are we to
remedy this problem? Implicitly, we have decoupled metric fluctuations
from our discussion so far, but now we are forced, by the asymmetric warping,
to reintroduce $\dd g$ fluctuations to satisfy the equations of
motion. Said differently, there is a
coupling of the form
\be\label{cross} \dd(\hat{*} G_4) \wedge  d\dd A_3 \ee
where $ \dd ( \hat{*} G_4)$ is a metric fluctuation. The background flux
read from \C{ranktwoM}\ has the form
$$ G_4 = B_{12} \wedge \omega_{b}.$$ This leads to a
non-vanishing coupling between $ \dd g$ and $ d\dd A_3$.

The second term in \C{expansion}\ is a $(5,3)$ form. Let us take the
case of $a=1$. To cancel these
terms by appropriate metric fluctuations, note that \C{cross}\
contributes the following terms to the $\dd A_3$ equation of motion:
\bea \label{metricfluct}
 & d \big(  \sqrt{g} \, \e_{\m_1 \cdots \m_6 2} \, dx^{\m_1} \cdots dx^{\m_5} 
\, \dd g^{\m_6 1} g^{22} \, B_{12}\, *\omega_b \, + \cr 
&  \sqrt{g} \, \e_{\m_1 \cdots 1 \m_7} \, dx^{\m_1} \cdots dx^{\m_5} 
\,  g^{11} \dd g^{\m_7 2} \, B_{12}\, * \omega_b \big). 
\eea
These terms give rise to both $(5,3)$ and $(6,2)$ forms. The first two $(5,3)$
terms of \C{metricfluct}\ can cancel the terms of \C{expansion}\ with
$a=1$ but not $a=2$ if
$$   \dd g_\m^{1} \, B_{12} \sim F_{2\m}, \qquad  \dd g_\m^{2} \, B_{12} \sim F_{1\m}. $$ 
The proportionality constant is a function of $r$. The $(6,2)$ terms
have the form $ d* F_2 \wedge \omega' $ for some internal space 2-form
$\omega'$. These terms modify the gauge-field kinetic terms. 

In a similar way, to cancel the terms
with $a=2$, we also need to consider $\dd g^{11}$ and $\dd g^{22}$
fluctuations. To determine the correct combination of metric
fluctuations, we should analyze the equation for metric
fluctuations which comes from terms in the action with the schematic form,
\be \int \dd g \Delta_L  \dd g + \dd(\hat{*} G_4) \wedge \dd(\hat{*}
G_4) + \dd(\hat{*} G_4) \wedge  d\dd A_3, \ee
where $\Delta_L$ is the operator obtained by expanding $\sqrt{g} R$ to
quadratic order in the metric fluctuations. 
With a suitable gauge choice, $ \Delta_L$ is the Lichnerowicz
Laplacian. The last two terms arise because of the non-vanishing
background $G_4$. 

Metric fluctuations are not the only fluctuations induced by our
original choice \C{fluct}. The $(2,2)$ form $d\dd A_3$ plugged into
the right hand side of \C{eomC}\ gives a $(4,4)$ form. To cancel this
term, we need to supplement the original $\dd A_3$ with an induced term of the
form
\be \label{indthree} \dd A_3' \sim *(B_{12} \wedge F_2) \, \zeta, \ee
where $\zeta$ is a 0-form on the internal space.
This $(3,0)$ form is chosen so that the $(4,4)$ pieces of 
$$  d\,\hat{*}\, d\, \dd A_3' = - G_4 \wedge d \dd A_3 $$
cancel up to a possible $(5,3)$ term on the left hand side. 
In turn, this new induced fluctuation $\dd A_3'$
mixes, via the Chern-Simons term and the 4-form kinetic term, with
$\dd A_3$ and $\dd g$. 
Rather than continue along this path, which is quite
involved, let us turn to an
alternate method based on duality.

\subsection{Insights from duality chasing}

Let us recall the method that we used to generate the warped
supergravity solutions. 
We started with the 11-dimensional Taub-NUT solution.  After reducing
along the circle direction to
get a IIA D6-brane, we T-dualized to get a smeared D5-brane of type IIB.  We then
performed a change of coordinates, and compactified a new
direction. In the static case, we effectively T-dualized back to type
IIA on a non-rectangular torus generating a $B_2$-field. In the
time-dependent case, we switched to coordinates in which the
null-brane quotient is a simple circle identification. Again, we T-dualized
back to IIA. Generating higher rank $B_2$-fields simply required more
T-dualities. In either case, the result is a IIA D6-brane with
$B_2$-fields along some world-volume directions. Finally, we lifted
these configurations to $11$ dimensions. 

At the starting point of this duality chain, the correspondence
between 11-dimensional supergravity and the low-energy theory on the
brane world-volume is well understood.  As discussed in section
\ref{prelims}, the gauge fluctuations on the
brane correspond to $3$-form fluctuations of the form $\dd C_3 = A_1\w\mo$,
where $A_1$ becomes the gauge-field on the brane, and $\mo$ is the
normalizable harmonic $2$-form on Taub-NUT. The overall normalization of
$\mo$ is determined by comparing the membrane and open string
actions. On reduction, the correct Yang-Mills
coupling and action emerge.   

To take advantage of our understanding in the basic case, and to learn more
about the warped backgrounds, it is natural to take the
known localized fluctuations from Taub-NUT and push them through the chain of
dualities.  By construction, we should obtain fluctuations of the
$11$-dimensional fields which are localized on the brane, and which
solve the equations of
motion (so that they give rise to massless $7$-dimensional fields).
Let us perform this exercise first for the static case with a rank $2$
$B_2$-field, and then for the time-dependent case.

\subsubsection{Duality chasing the static background}
\label{staticchase}
We start with the ordinary Taub-NUT gauge fluctuations studied earlier,
\be
\dd A_3 = \left(A_\m dx^\m + A_5 dx^5 + A_6 dx^6 \right)\w\mo
\ee
where
\be
\mo = 2\pi\ell_p^2 \,d\left[H_1^{-1} \left(dy + \frac{R}{2\cos\al} \cos\T 
d\psi\right)\right].
\ee
The warp factor $H_1$ here, and $H_2$ appearing below, are defined in section 
\ref{twisted}.  Note that $R$ has been rescaled to $R/\cos\al$ to agree with
the twisted solution.  This is the same procedure that was followed in section
\ref{twisted}, but this time we will carry the fluctuations with us under the 
successive dualities.

After chasing this fluctuation through the duality chain, we find that the $A_\m$ fluctuations
appear as $3$-form fluctuations in the new background:
\bea \label{chasefluct}
\dd A_3 &=& 2\pi\ell_p^2 A_\m dx^\m \w d\left[H_1^{-1}\left(dy + 
\hlf R\cos\T d\psi\right)\right] \non\\
&& - 2\pi\ell_p^2 \sin\al H_1^{-1} \left(\ast_5 F_2\right). 
\eea
In the second line, the $\ast_5$ acts only in the $\m$, $\n$
directions.  Explicitly in components 
\be
\dd A_{\m\n\rho} = - 2\pi\ell_p^2 \sin\al H_1^{-1} \e_{\m\n\rho}
\vphantom{\e}^{\lambda\s} \p_\lambda A_\s. 
\ee
This induced $3$-form fluctuation had to have been there to cancel the
contribution to the equation of motion from the Chern-Simons
coupling. This is precisely the induced $(3,0)$ form $\dd A_3'$
described in \C{indthree}. However, here we have an explicit form for
the fluctuation. We can indeed check that \C{chasefluct}\  satisfies
\be
d \,\hat{\ast}\, d\, \dd A_3 =  G_4 \w\dd A_3. 
\ee

The $A_5$ and $A_6$ fluctuations give rise to both a $3$-form
fluctuation, 
and metric fluctuations, in a particular combination.  The $3$-form
component is given by
\bea
\label{dchasea}
\dd A_3 &=& 2\pi\ell_p^2 \left(A_5 dz_1 + A_6 dz_2 \right)\w\left[
\frac{R}{2r^2}H_1^{-1} H_2^{-1} dr\w\left(dy + \hlf R\cos\T d\psi\right)\right.
\non\\
&& \left. - \frac{R}{2\cos\al}H_1^{-1} \sin\T d\T\w d\psi\right]
\eea
while the metric fluctuations are most simply written as a pair of vielbein
fluctuations,
\bea
e^{z_1} &=& H_1^{\frac{1}{3}} H_2^{-\frac{1}{3}} \left[dz_1 + 2\pi\ell_p^2 
\frac{R}{2r^2} H_1^{-2} \tan\al A_6 dr\right] \non\\
e^{z_2} &=& H_1^{\frac{1}{3}} H_2^{-\frac{1}{3}} \left[dz_2 - 2\pi\ell_p^2 
\frac{R}{2r^2} H_1^{-2} \tan\al A_5 dr\right].
\eea
That a combination of metric and $3$-form fluctuations are needed is
in accord with our earlier direct analysis.

Suppose we consider a different set of fluctuations that differ by a gauge
transformation $${A_\m}' = A_\m + \p_\m \lambda, \qquad {A_i}' = A_i + \p_i \lambda.$$  
Before chasing this fluctuation through the duality chain,
we know that this corresponds to the same supergravity solution because it
differs from our original configuration 
by a 3-form gauge transformation 
$$A_3 \rr A_3 + d(\lambda\mo).$$  After
performing the dualities, we must therefore also have the same
solution. However, now even the metric
differs for $A$ and $A'$.  The resolution must be that the two answers
differ by some combination of 3-form shift and diffeomorphism.  In this
way, what we would have thought of as a $U(1)$ gauge symmetry becomes
mixed with diffeomorphisms, and in this way, the resulting theory can be
reinterpreted as having a non-commutative gauge group.

With some foresight, let us define $A_1 = A_5 \cos\al$, $A_2 = A_6 \cos\al$.
The latter of these two redefinitions is natural from the change of variables 
(\ref{xtoz}) between $x^6$ and $z^2$.  The definition of $A_1$ can then be
justified from the symmetry between $z^1$ and $z^2$.  The two equations 
above may now be written
\be
e^{z_i} = H_1^{\frac{1}{3}} H_2^{-\frac{1}{3}} \left[dz^i + 2\pi\ell_p^2 
\tan\al\, d\left(H_1^{-1} \right) \e^{ij} A_j \right].
\ee
These results are exact, at least to the extent that supergravity can be 
trusted at each step in the duality chain.  However, in our subsequent
discussion in this section, we will
work only to linear order in the gauge fluctuation $A$.

The form of the metric fluctuation suggests a natural change of coordinates
\be
Z^i = z^i + 2\pi\ell_p^2 \tan\al H_1^{-1} \e^{ij} A_j
\ee
where $i=1,2$. 
This diffeomorphism moves all the metric fluctuations into the 
world-volume; in other words, only components of the metric that have no
Taub-NUT indices fluctuate. However, there is no unique choice of
diffeomorphism. There are other diffeomorphisms that
accomplish the same task since only
the $r$-dependence is fixed by this constraint.  

Specifically, let us 
consider a more general change of coordinates
\be
\label{thetadiffeo}
Z^i = z^i + 2\pi\ell_p^2 \left(\tan\al H_1^{-1} \e^{ij} - \T^{ij} 
\right)A_j
\ee
where $\T^{ij}$ is a constant anti-symmetric matrix.  The suggestive
label is no mere
coincidence.  As we shall see, this theta will have an interpretation as the 
non-commutativity parameter of the world-volume theory.  Of course, we
could consider more general 
diffeomorphisms, but it turns out that these particular ones are especially nice.
Also, when this diffeomorphism, \C{thetadiffeo}, is pulled back to the brane
world-volume, 
i.e. computed at $r=0$, then the first term drops out leaving simply
\be
Z^i = z^i - 2\pi\ell_p^2 \T^{ij} A_j.
\ee
Under (\ref{thetadiffeo}), the metric fluctuations become
\be
e^{z_i} = H_1^{\frac{1}{3}} H_2^{-\frac{1}{3}} \left[dZ^i - 2\pi\ell_p^2 \left(
\tan\al H_1^{-1} \e^{ij} - \T^{ij} \right) \left(\p_\m A_j dx^\m + 
\p_k A_j dZ^k \right)\right]. 
\ee
The background $3$-form,
\be
\label{abackgr}
\langle A_3 \rangle = H_2^{-1} \tan\al dz_1 \w dz_2 \w \left(dy + \hlf R\cos\T
d\psi\right)
\ee
also changes.  The combination of fluctuating $3$-forms becomes, to 
linear order in $A$ (also neglecting the induced $3$-form from the second line
of \C{chasefluct}),
\bea
\dd A_3 &=& 2\pi\ell_p^2 \left(A_\m dx^\m + \frac{1}{\cos^2 \al} A_i dZ^i 
\right)\w d\left[H_1^{-1}\left(dy + \hlf R\cos\T d\psi\right)\right] \non\\
&& - 2\pi\ell_p^2 \tan\al H_2^{-1}\left(\tan\al H_1^{-1} - \T^{12} 
\right)\left(\p_\m A_i dx^\m + \p_j A_i dZ^j \right)\w dZ^i \non\\
&& \w \left(dy + \hlf R\cos\T d\psi\right). 
\eea
This expression can be cleaned up by making a $3$-form gauge transformation.
Specifically by adding an exact $3$-form
\be
d\left[2\pi\ell_p^2 \tan\al H_2^{-1} \left(\tan\al H_1^{-1} - \T^{12}
\right) A_i dZ^i \w \left(dy + \hlf R\cos\T d\psi\right)\right], 
\ee
we obtain the total fluctuating $3$-form
\bea
\dd A_3 &=& 2\pi\ell_p^2 A_\m dx^\m \w d\left[H_1^{-1}\left(dy + \hlf R\cos\T 
d\psi\right)\right] \non\\
&& + 2\pi\ell_p^2 A_i dZ^i \w d\left[\left(1 + \tan\al\T^{12} \right)
H_2^{-1} \left(dy + \hlf R\cos\T d\psi\right)\right]. 
\eea

As mentioned earlier, we can choose any constant value for the
parameter $\T^{12}$.  Let us consider three particular choices of $\T^{12}$
that simplify the above fluctuations.

The first choice we consider is $\T^{12} = 0$.  For this choice, the
diffeomorphism (\ref{thetadiffeo}) vanishes at $r=0$.  As we shall see, this 
means effectively that on the brane, we see only commutative gauge 
transformations.  However, we cannot really escape non-commutativity
in the full $11$-dimensional
theory in the sense that a commutative gauge transformation on $A$
still maps to a diffeomorphism. 

The next choice is $\T^{12} = -1/ \tan\al$, i.e. 
$\T = B^{-1}$.  In this case, the $3$-form piece above vanishes and the $A_i$ 
fluctuations move entirely into the metric.
The metric fluctuations become explicitly
\be
e^{z_i} = H_1^{\frac{1}{3}} H_2^{-\frac{1}{3}} \left[dZ^i - 2\pi\ell_p^2
\frac{1}{\sin\al\cos\al} H_1^{-1} H_2 \e^{ij} \left(\p_\m A_j dx^\m + 
\p_k A_j dZ^k \right)\right].
\ee
At $r=0$ this reduces to
\be
e^{z_i} = H_1^{\frac{1}{3}} H_2^{-\frac{1}{3}} \left[dZ^i - 2\pi\ell_p^2
\frac{1}{\tan\al} \e^{ij} \left(\p_\m A_j dx^\m + \p_k A_j dZ^k
\right)\right]. 
\ee

One more choice worth mentioning is 
$\T^{12} = -\sin\al\cos\al$.  If we believe the correspondence between the
$\T$ appearing in the diffeomorphism and the $\T$ of non-commutative 
Yang-Mills, then this should correspond to pure NCYM.  We find for this
choice that the $3$-form fluctuation becomes
\bea
\dd A_3 &=& 2\pi\ell_p^2 A_\m dx^\m \w d\left[H_1^{-1} 
\left(dy + \hlf R\cos\T d\psi\right)\right] \\
&& + 2\pi\ell_p^2 A_i dZ^i \w d\left[\cos^2 \al H_2^{-1} 
\left(dy + \hlf R\cos\T d\psi\right)\right]. \non
\eea
Near $r=0$, $H_1^{-1} \simeq \cos^2 \al H_2^{-1}$, so $A_\m$ and $A_i$
appear on equal footing.  This is precisely what we would expect for pure
NCYM.

Finally, let us consider what happens to the commutative gauge group of our
starting point.  Here, we take the point of view that we have fixed
a diffeomorphism initially. The duality chain and the diffeomorphism
then define  a map from the
original theory of supergravity on a Taub-NUT space to a new theory
with flux.  Under a gauge transformation $A \rr A + d\lambda$, 
we are then instructed to perform a further diffeomorphism of the form 
(\ref{thetadiffeo}) but with $A_j$ replaced by $\p_j \lambda$ (in addition to
simply shifting $A_\m$ and $A_i$ by $\p_\m \lambda$ and $\p_i \lambda$).  This
diffeomorphism acts non-trivially on all of the gauge fields, and on any other
fields that we might consider, such as fields that correspond to scalars
on the D6-brane.  Explicitly, at $r=0$, a field would transform as
\be
\dd \Phi = 2\pi\ell_p^2 \T^{ij} \p_j \lambda \p_i \Phi. 
\ee
So, in total, a gauge field like $A_\m$ would transform as
\be
\dd A_\m = \p_\m \lambda + 2\pi\ell_p^2 \T^{ij} \p_j \lambda \p_i
A_\m. 
\ee
These are, of course, the expected noncommutative gauge transformations
to linear order in $\T$.

\subsubsection{Relation to the Seiberg-Witten map}
\label{swmap}
In this section we will set $2\pi\ell_p^2 = 1$.
The diffeomorphism (\ref{thetadiffeo}) for the case $\T=B^{-1}$ has an
interesting relation to the Seiberg-Witten map relating the
non-commutative gauge field $\hat{A}$ to the ordinary gauge
field $A$\cite{Seiberg:1999vs}. As shown in
\cite{Cornalba:1999ah}, in the presence of
a background $B_2$-field, we can define (for $\T=B^{-1}$) the following
diffeomorphism on the world-volume of a single D-brane:
\be
\label{cornadiffeo}
X^i = x^i + \T^{ij}\hat{A}_j(x).
\ee
Under this diffeomorphism,
\be
\label{cornafandb}
(F_{ij}(X) + B_{ij})\frac{\p X^i}{\p x^k}\frac{\p X^j}{\p x^l} = B_{kl}
\ee
so that the coordinates $x$ are interpreted as the coordinates in which
the commutative field strength $F$ is constant.
This diffeomorphism therefore moves the gauge
fluctuations on the brane entirely into the metric.
Moreover, the
diffeomorphism (\ref{cornadiffeo}) is not unique, but only defined up to
diffeomorphisms that leave $B$ invariant.

We have already seen that under the diffeomorphism (\ref{thetadiffeo}) for
$\T=B^{-1}$, the gauge field fluctuations are moved entirely into the
metric, at least to linear order in $A$. Based on its remarkable similarity to
(\ref{cornadiffeo}), it is natural to wonder whether we can move 
the gauge fluctuations into the metric to all orders in $A$.
We will argue below that this is indeed the case provided
we replace $A$ in (\ref{thetadiffeo}) by $\hat{A}$. Specifically, we will
use the diffeomorphism
\be
\label{newthetadiff}
Z^i = z^i + \left(\tan\al H_1^{-1} \e^{ij} - (B^{-1})^{ij} 
\right)\hat{A}_j(Z)
= z^i +
\frac{1}{\sin\al\cos\al} H_1^{-1} H_2 \e^{ij} \hat{A}_j(Z).
\ee
To facilitate comparison with (\ref{cornadiffeo}), 
it is convenient to rewrite (\ref{newthetadiff}) as 
\be
\label{newthetadiffeo}
z^i = Z^i - \frac{1}{\sin\al\cos\al} H_1^{-1} H_2 \e^{ij} \hat{A}_j(Z).
\ee
The coordinate $z$ is thus the analogue of $X$ in (\ref{cornadiffeo}),
while $Z$ is the analogue of $x$.
Instead of applying this diffeomorphism to the fields
$\langle A_3\rangle$ and $ \dd A_3 $ respectively, it is more
useful to apply it to the field strengths
$\langle G_4 \rangle= d\langle A_3 \rangle$ and $\dd G_4 = d \dd A_3$.
Using the explicit forms of the background field (\ref{abackgr}) and
the fluctuation (\ref{dchasea}) obtained by the duality chasing, we compute
\bea
\label{Gbandf}
\langle G_4 \rangle &=& \tan\al\ dz^1\wedge dz^2 \wedge d[H_2^{-1} (dy + 
\frac{R}{2} \cos\T\ d\psi)], \\
\dd G_4 &=& \sec\al\ \left(F_{12} dz^1\wedge dz^2 + \p_\m A_i dx^\m \w dz^i 
\right) \non\\
&& \wedge \left[\frac{R}{2r^2}H_1^{-1}H_2^{-1} dr\wedge (dy +\frac{R}{2} 
\cos\T\ d\psi) - \frac{R}{2\cos\al}H_1^{-1} \sin\T\ d\T\wedge d\psi\right]
\non \\
&&+ \sec\al\tan^2\al \frac{R^2}{4r^2} H_1^{-2} H_2^{-1} \sin\T A_idz^i\w dr\w 
d\T\w d\psi
\eea
where we have only shown the terms in $\dd G_4$ that are independent of
$A_\mu$.
We now apply the diffeomorphism (\ref{newthetadiffeo}) to 
$\langle G_4 \rangle +\dd G_4$. 
In order to compare the result with (\ref{cornafandb}), which strictly
speaking, is valid for maximal rank $B_2$-field, we will now ignore all
dependence of the gauge fields on the coordinates $x^{\mu}$ which
correspond to worldvolume coordinates transverse to the $B$-field.
Having thus dropped all terms involving $\p_\mu A$, we find
to lowest order, $\hat A_i = A_i$ and
\be
\label{diffeoimagea}
\langle G_4\rangle +\dd G_4\mapsto
\tan\al dZ^1\wedge dZ^2 \wedge d\left[H_2^{-1} (dy + 
\frac{R}{2} \cos\T\ d\psi)\right].
\ee
Thus, to lowest order,  we can indeed interpret the $Z^i$ as the
coordinates in which the 4-form
field strength $G$ is constant (in $Z$), and~(\ref{diffeoimagea}) is the
natural generalization of~(\ref{cornafandb}). 

We shall now verify the second order correction, so we
write $\hat A_i = A_i + a_i$.  The second order piece, $a_i$, is a
function of the space-time coordinates, and of $r$, and is explicitly
given~by:
\be
\label{newswmap}
\hat A_i = A_i + a_i =
A_i + \hlf\tilde\T^{jk} \left(2A_k \p_j A_i - A_k \p_i A_j \right)
+ {\cal O}(A^3)
\ee
where we have defined the natural $r$-dependent combination
\be
\label{tilthetadef}
\tilde\T^{ij} = - \frac{1}{\sin\al\cos\al} H_1^{-1} H_2 \e^{ij}.
\ee
Note that this is precisely the tensor contracted with $\hat{A}_j$ 
in~(\ref{newthetadiffeo}).
To this order, we now find
\be
\label{diffeoimage}
\langle G_4\rangle +\dd G_4\mapsto
\tan\al dZ^1\wedge dZ^2 \wedge d\left[H_2^{-1} (dy + 
\frac{R}{2} \cos\T\ d\psi)\right] + P_i dZ^i \w dr
\w d\T\w d\psi,
\ee
where,
\bea
P_1 &=& -\frac{R\sin\T}{2\cos^2 \al}
\Big[\frac{R\sin^2 \al}{2r^2 \cos\al} H_1^{-2} H_2^{-1}
a_1 + H_1^{-1} \p_r a_1 \non \\
& & - \frac{R\sin\al}{2r^2 \cos^2 \al} H_1^{-3} \left(
2A_1 \p_2 A_1 - A_2 \p_1 A_1 - A_1 \p_1 A_2 \right)+ {\cal O}(A^3)\Big],
\eea
and similarly for $P_2$. In writing the above expression, we have used
\be
A_1(z) = A_1(Z) - \frac{1}{\sin\al\cos\al} H_1^{-1} H_2 \left(A_2 \p_1 A_1 - 
A_1 \p_2 A_1 \right) + {\cal O}(A^3).
\ee
We see from~(\ref{diffeoimage}) that we can interpret the $Z^i$ as the
coordinates in which the 4-form
field strength $G$ is constant (in $Z$) only if the $P_i$ vanish.
Due to the presence of the derivatives in $r$, the constraint $P_i=0$
provides a non-trivial
consistency check of the diffeomorphism~(\ref{newthetadiffeo})
with the choice of $r$-dependent noncommutativity
parameter~(\ref{tilthetadef}). 
Happily, the choice of $a_i$ given
by~(\ref{newswmap}) indeed satisfies this constraint.

At $r=0$, $\tilde\T^{ij}$ reduces to the previous $\T^{ij}$,
and~(\ref{newswmap}) reproduces
the Seiberg-Witten map between $A$ and $\hat A$ to linear
order in $\T$. In this sense
we can view equation~(\ref{newswmap}) as a lift of the usual
Seiberg-Witten map~\cite{Seiberg:1999vs} between
commutative and non-commutative variables to the full supergravity theory.
The diffeomorphism~(\ref{newthetadiffeo}) is the corresponding lift 
of equation~(\ref{cornadiffeo}).
It is very reasonable to expect that the above procedure can be iterated,
and that this lift can be constructed to all finite orders in $A$ (or
equivalently to all finite orders in $\T$).  We expect that all of
the statements of the previous section should similarly extend to
all orders.

\subsubsection{Duality chasing the time-dependent background}

Before jumping into a time-dependent duality chase, it is worth surveying the
structure of the time-dependent solution \C{11Dtimedep}, which takes the form:
\bea \label{survey}
ds^2 &=& f_1(r, x^+) \left(-2dx^+ dx^- + dx_3^2 + dx_4^2
+ dx_5^2 \right) + f_2(r,x^+) \left(dx^2 + dz^2 \right)
\\
&&  + f_3(r,x^+) \left(-x^2 (dx^+)^2 +
2x^+ xdx^+ dx\right)+ f_4(r,x^+) ds_{\M_1}^2 + f_5(r,x^+) ds_{\M_2}^2. \non
\eea
The accompanying M theory $G_4$ has both $(3,1)$ and $(2,2)$
pieces. There are multiple warp factors in the metric \C{survey}\ but
what is most remarkable is that the internal space warp factors depend
on $x^+$. In this sense, space-time and the internal space are warped
with respect to one another. A direct analysis of small fluctuations, along
the lines of section~\ref{staticwarped}, should demonstrate the
existence of a localized gauge-field in an interesting way. 
However, the same issues that we met in the static case
will also appear in this case. We will therefore follow the duality
chasing tactic again. 

Let us start again with the fluctuation $A_1\w\mo$, where 
\be
\mo = 2\pi\ell_p^2 \,d\left[H^{-1} \left(dy + \hlf R\cos\T
  d\psi\right)\right]. 
\ee
On T-dualizing $z$, changing to invariant null-brane coordinates, T-dualizing
back, and lifting to $11$ dimensions, we find again a mixing of $3$-form
and metric fluctuations.  Let us first define the following combinations
of the original gauge fluctuations which naturally appear at the end of the
duality chain:
\bea
\tilde A_+ &=& A_+ + \tilde z A_x + \hlf\tilde z^2 A_- \non\\
\tilde A_x &=& A_x + \tilde z A_-.
\eea
We drop the tildes from now on.  Furthermore, we make the same definitions as 
before so that the coordinates agree with (\ref{11Dtimedep}).  Then the result 
is a $3$-form fluctuation
\bea
\dd A_3 &=& 2\pi\ell_p^2 \left[A_i dx^i + A_- dx^- + Hh^{-1} A_z dz
+ \left(A_+ + h^{-1}\left(\frac{x^+ x}{R^2} A_x + \frac{x^2}{R^2} A_- \right)
\right)dx^+ \right.\non\\
&& \left. + \left(Hh^{-1} A_x - h^{-1} \frac{x^+ x}{R^2} A_- \right) dx
\right]\w\left[\frac{c}{r^2} H^{-2} dr\w\left(dy + c\cdot\cos\T d\psi\right)
\right] \\
&& - cH^{-1} \sin\T \left(A_+ dx^+ + A_- dx^- + A_x dx + A_z dz + A_i
dx^i \right)\w d\T\w d\psi. \non
\eea
The metric fluctuations can be most succinctly written by making some 
replacements in the metric (\ref{11Dtimedep})
\bea
& dx^- &\rr dx^- + 2\pi\ell_p^2 \frac{c}{r^2} H^{-2} \frac{x}{R} A_z dr \non\\
& dx &\rr dx + 2\pi\ell_p^2 \frac{c}{r^2} H^{-2} \frac{x^+}{R} A_z dr \\
& dz &\rr dz - 2\pi\ell_p^2 \frac{c}{r^2} H^{-2} \left(\frac{x^+}{R} A_x
+ \frac{x}{R} A_- \right) dr. \non
\eea

As before, a particular change of coordinates presents itself for use
in moving the metric fluctuations purely into the world-volume 
directions.  If we let $\m$ run over the indices ($-$, $x$, $z$), then a
natural diffeomorphism is
\be
X^\m = x^\m - 2\pi\ell_p^2 H^{-1} \T^{\m\n} A_\n
\ee
where, 
\bea
\T^{z-} &=& -\T^{-z} = \frac{x}{R} \\
\T^{zx} &=& -\T^{xz} = \frac{x^+}{R}, \non
\eea
with all other components of $\T$ zero.  This agrees with the results
reported in~\cite{Hashimoto:2002nr}.

We close by noting that under this diffeomorphism, the fluctuating
part of the $3$-form becomes (to
linear order in $A$)
\bea
\dd A_3 &=& 2\pi\ell_p^2 \left(A_+ dx^+ + A_i dx^i + A_\m dx^\m \right)\w
d\left[H^{-1} \left(dy + c\cdot\cos\T d\psi\right)\right] \\
&& - 2\pi\ell_p^2 H^{-1} h^{-1} \left[\left(\frac{x^+}{R}\right)^2 d\left(
A_z dz\right) + \left(\frac{x}{R^2} A_- + \frac{x^+}{R^2} A_x \right) dx^+ \w
dx \right.\non\\
&& \left. + \left(\frac{x}{R} dx^+ - \frac{x^+}{R} dx\right)\w\left(
\frac{x^+}{R} dA_x + \frac{x}{R} dA_- \right)\right]\w\left(dy + c\cdot\cos\T 
d\psi\right).  \non
\eea
Clearly, there is much more to be said. This analysis can be continued
along the lines of
section~\ref{staticchase}\ and section~\ref{swmap}, perhaps giving an
analogue from gravity of the Seiberg-Witten map for the time-dependent case.

\section*{Acknowledgements}
The work of K.~D. is supported in part by a David and Lucile Packard
Foundation Fellowship 2000-13856. The work of G.~R. is supported in
part by DOE Grant DE-FG02-90ER-40560 and by NSF CAREER Grant No.
PHY-0094328, while the work of D.~R. is supported
in part by a Julie Payette--NSERC PGS B Research Scholarship.
The work of S.~S. is supported in part by NSF CAREER Grant No.
PHY-0094328, and by the Alfred P. Sloan Foundation.

\appendix
\section{Conventions and Rules}
\label{conventions}
In performing type II T-duality and lifts to M theory, we will
(mostly) use the conventions
found in \cite{Johnson:2000ch}. Let $\ell_p$ denote the
eleven-dimensional Planck scale, and $\ell_s$ denote the string scale. 

\subsection{Lifting IIA to M theory}

A type IIA SUGRA configuration can be lifted to 11 dimensions.  Let
$y$ parametrize the eleventh direction. The $y$
coordinate is compact with periodicity $2\pi$.  Writing the 11-dimensional
fields on the left and the IIA fields on the right, we have:

\bea
G_{\m\n} &=& e^{-\frac{2}{3}\Phi}G_{\m\n}^{(IIA)} + 
e^{\frac{4}{3}\Phi}C_\m C_\n \non\\
G_{\m y} &=& -\ell_p e^{\frac{4}{3}\Phi}C_\m \non\\
G_{yy} &=& \ell_p^2 e^{\frac{4}{3}\Phi} \\
A_{\m\n\rho} &=& C_{\m\n\rho} \non\\
A_{\m\n y} &=& \ell_p B_{\m\n}. \non
\eea

\subsection{T-Duality}

We dualize in the direction $x$, which we scale to have periodicity $2\pi$
(and is hence dimensionless).

The NS-NS fields transform in the following manner:
\bea
e^{2\Phi'} &=& \frac{\ell_s^2 e^{2\Phi}}{G_{xx}} \non\\
G_{xx}' &=& \frac{\ell_s^4}{G_{xx}} \non\\
G_{\m x}' &=& \frac{\ell_s^2 B_{\m x}}{G_{xx}} \\
G_{\m\n}' &=& G_{\m\n} - \frac{G_{\m x}G_{\n x} - B_{\m x}B_{\n x}}{G_{xx}}
\non\\
B_{\m x}' &=& \frac{\ell_s^2 G_{\m x}}{G_{xx}} \non\\
B_{\m\n}' &=& B_{\m\n} - \frac{B_{\m x}G_{\n x} - G_{\m x}B_{\n x}}{G_{xx}}.
\non
\eea

The transformation of the R-R potentials is given by
\bea
\ell_s^{-1}{C_{\m\cdots\n\alpha x}^{(n)}}' &=& C_{\m\cdots\n\alpha}^{(n-1)} -
(n-1)\frac{C_{[\m\cdots\n|x}^{(n-1)}G_{|\alpha]x}}{G_{xx}} \\
\ell_s {C_{\m\cdots\n\alpha\beta}^{(n)}}' &=&
C_{\m\cdots\n\alpha\beta x}^{(n+1)} + nC_{[\m\cdots\n\alpha}^{(n-1)}B_{\beta]x}
+ n(n-1)\frac{C_{[\m\cdots\n|x}^{(n-1)}B_{|\alpha|x}G_{|\beta]x}}{G_{xx}}. \non
\eea


\providecommand{\href}[2]{#2}\begingroup\raggedright\endgroup

\end{document}